\documentstyle[12pt]{article}

\renewenvironment{thebibliography}[1]
    {\begin{list}{\arabic{enumi}.}
    {\usecounter{enumi}\setlength{\parsep}{0pt}
\setlength{\leftmargin 1.25cm}{\rightmargin 0pt}
     \setlength{\itemsep}{0pt} \settowidth
    {\labelwidth}{#1.}\sloppy}}{\end{list}}

\begin{document}
\def\ETslash{\not{\hbox{\kern-4pt $E_T$}}}
\def\mynot#1{\not{}{\mskip-3.5mu}#1 }
\def\sss{\scriptscriptstyle }
\def\ra{\rightarrow}
\def\d{{\rm d}}
\def\M {{\cal M}}   
\def\WW {W_L W_L}
\def\MWW {M_{WW} }
\def\WWWW {W_L W_L \ra W_L W_L}
\def\PPPP {\phi \phi \ra \phi \phi}
\def\WPWM{ W^+(\ra \ell^+\nu) W^-(\ra q_1 \bar q_2) }
\def\WPZ{ W^+(\ra \ell^+ \nu) Z^0(\ra q \bar q) }
\def\WPWP{ W^+(\ra \ell^+ \nu) W^+(\ra \ell^+ \nu) }
\def\etc{ {\it etc.}}
\def\be{\begin{equation}}
\def\ee{\end{equation}}
\def\bea{\begin{eqnarray}}
\def\eea{\end{eqnarray}}
\def\CPbar{\hbox{{\rm CP}\hskip-1.80em{/}}}
\def\D0{D\O~}
\def\pbarp{ \bar{{\rm p}} {\rm p} }
\def\pp{ {\rm p} {\rm p} }
\def\ifb{ {\rm fb}^{-1} }
\def\del{\partial }
\def\ra{\rightarrow}
\setcounter{footnote}{1}
\renewcommand{\thefootnote}{\fnsymbol{footnote}}
%=========================Title Page===============================

\begin{titlepage}

%%%%%%%%%%%%%%%%%%%%%%%%%%%%%%%%%%%%%%%%%%%%%%%%%%%%%
%hep-ph/97XXXXX \hfill {\small MSUHEP-70316} 
%\vspace{-0.3cm}

\begin{flushright}
{\small hep-ph/9712513}\\
{\small MSUHEP-70316}\\
{ March 1997}
\end{flushright}
%%%%%%%%%%%%%%%%%%%%%%%%%%%%%%%%%%%%%%%%%%%%%%%%%%%%%
\vspace*{1.2cm} 
\centerline{\large\bf  
Proposals for Studying TeV $\WWWW$}
\baselineskip=18pt
\centerline{\large\bf  
Interactions Experimentally\footnote{
To be published in ``Perspectives On Higgs Physics'', 
second edition, 
edited by Gordon L. Kane, World Scientific, Singapore.} }
\vspace*{1.2cm}
\baselineskip=17pt
\centerline{\normalsize  
 {\bf C.--P. Yuan} } 
\vspace*{0.4cm}
\centerline{\normalsize\it
Department of Physics and Astronomy, Michigan State University }
\centerline{\normalsize\it
East Lansing, Michigan 48824 , USA }

\vspace{0.5cm}
\begin{abstract}
\noindent
We discuss how to experimentally study the electroweak symmetry 
breaking (EWSB) sector by observing
$\WWWW$ interactions in the TeV region. 
We discuss some general features of the event structure in the 
signal and the background events. Various techniques 
to enhance the signal-to-background ratio 
are also presented. We show how to detect longitudinal $W$-bosons
either in the central rapidity region of the detector or in 
the beam pipe direction.
Finally, we globally classify the sensitivities of 
the Large Hadron Collider and the Linear Colliders to probing
the next-to-leading order bosonic operators 
of the electroweak chiral lagrangian used to parametrize 
models of strongly coupled EWSB sector.

\end{abstract}

\end{titlepage}

%%%%%%%%%%%%%%%%%%%%%%%%%%%%%%%%%%%%%%%%%%%%%%%%%%%%%%%%%%%%
% 1. Introduction
% 2. Signal
% 3. Backgrounds
% 4. How to distinguish signal from background 
% 5. Various models 
% 6. Sensitivities of high energy colliders to EWSB sector
% 7. Discussions and Conclusions
%%%%%%%%%%%%%%%%%%%%%%%%%%%%%%%%%%%%%%%%%%%%%%%%%%%%%%%%%%%%

\newpage
\renewcommand{\thefootnote}{\arabic{footnote}}
\setcounter{footnote}{0}
\baselineskip=18pt 

%=========================Body of paper=======================

\section{ Introduction }
\indent\indent

The current low energy data are already sensitive to the 
$SU(2)_L\times U(1)_Y$ gauge interactions of the Standard Model (SM), 
but not  yet sensitive to
the electroweak symmetry breaking (EWSB) mechanism of the SM,
i.e. the Higgs mechanism.
Neither do we know anything
about the origin of the fermion mass which is generated 
in the SM by the Yukawa interactions among the fermions and the Higgs boson.
The existence of
light resonance(s) originating from the EWSB sector with mass(es) 
well below the TeV scale is a possibility in the SM, and 
a necessity in its supersymmetric extensions 
(i.e. supersymmetry models).
In both cases,  these particles shall be detected
at the high energy colliders such as the CERN Large Hadron Collider (LHC),
a proton-proton collider, 
and the future electron (and photon) Linear Colliders (LC)~\cite{lcws93}.
However, the Higgs boson can be very heavy ($\sim 1$\,TeV), subject to
the triviality and the unitarity bounds \cite{light}, and its mass can 
merely serve as a cutoff at the TeV scale beyond which new physics 
must show up.
If the EWSB is indeed driven by strong interactions and there is no new 
resonance well below the TeV scale,  then the interactions among the
longitudinal $W$'s must become strong in the TeV region.
How to experimentally probe the strongly coupled 
EWSB sector is the subject of this article.
Since for models with light resonance(s) in the symmetry breaking sector,
the interactions among the Goldstone bosons in the TeV region 
cannot become strong,\footnote{
The potentially bad high energy behavior of 
the scattering matrix element (if without any light resonance)
is cut off by the tail of the light resonance.
}
we shall not consider that class of models here.

In the spontaneous symmetry breaking sector, 
the would-be Goldstone bosons ($\phi$'s) characterize 
the broken symmetry of the theory, and become the longitudinal degree of 
freedom of the massive $W$-bosons.
Consequently, a study of the symmetry breaking sector 
requires an understanding 
of the interactions of these would-be Goldstone bosons. 
According to the electroweak equivalence theorem (ET)
\cite{et1,et2,et3},
the S-matrix of $\WWWW$ is the same as 
that of $\PPPP$ in the limit of $E_W \gg M_W$,\footnote{
We shall use $W$ to denote either $W^\pm$ or $Z^0$,
unless specified otherwise.} where
$E_W$ is the energy of the $W$-boson in the center-of-mass frame of the
$WW$ pair, and $M_W$ is the mass of the $W$-boson.
Hence, to probe the EWSB sector, 
it is necessary to detect the longitudinal 
$W$ pairs produced via the $\WW$ fusion mechanism
\cite{effw} in the TeV region.
Below, we will show how to detect the $\WWWW$ signal. 

In section 2, we discuss the possible signals 
predicted by various models of EWSB mechanism.
In section 3, we discuss the large backgrounds involved in the 
detection of the signals.
In section 4, we discuss the characteristic differences between the event 
structures of the signal and the background. In section 5, 
we give some recipes for detecting the signal predicted by various models.
In section 6, we discuss the
sensitivities of future high energy colliders to probing
the EWSB sector.
Section 7 contains our discussions and conclusions.

\section{ Signal }
\indent\indent

The event signature of the signal is a longitudinal $W$-pair produced 
in the final state. 
Assuming no light resonance(s) \cite{light}, 
the electroweak symmetry breaking sector in the TeV region
may either contain a scalar- or vector-resonance, \etc, 
or no resonance at all.
For a model with a TeV scalar (spin-0,isospin-0) resonance, 
the most useful detection
modes  are the $W^+W^-$ and $Z^0Z^0$ modes
which contain large isospin-0 channel contributions.
For a model with a TeV vector (spin-1,isospin-1) resonance, the most
useful mode is the $W^\pm Z^0$ mode because it contains 
a large isospin-1 channel contribution.
If there is no resonance present in the symmetry breaking sector, all the 
$WW$ modes are equally important, so the $W^\pm W^\pm$ mode is also 
useful. Actually, because of the small SM backgrounds for the 
$W^\pm W^\pm$ mode, this can become the most important 
detection mode if no TeV resonance is present in the EWSB sector. 

Before we discuss the backgrounds, we have to specify the decay mode of 
the $W$-bosons in the final state. 
Let's first concentrate on the cleanest final state, i.e. the pure
leptonic decay mode. 
The branching ratios of $W^+ \ra \ell^+ \nu$ and 
$Z^0 \ra \ell^+\ell^-$ are 2/9 and 0.06, respectively, for
$\ell^+=e^+$ or $\mu^+$.
If the $Z^0Z^0$ pair signal is large enough, the 
$Z^0(\ra \ell^+\ell^-)Z^0(\ra \ell^+\ell^-)$
and $Z^0(\ra \ell^+\ell^-)Z^0(\ra \nu \bar \nu)$ modes would be most
useful at hadron colliders \cite{ZZllnn}.
Otherwise, it is also necessary to include the 
$W^\pm W^\mp$, $W^\pm Z^0$ and $W^\pm W^\pm$ modes \cite{llll}.
Although the pure leptonic mode gives the cleanest signal signature,
its event rate is small 
because of the relatively small branching ratio.
To improve the signal event rate for discriminating models of EWSB, 
one should also study the other decay modes, such as the lepton plus jet 
modes at the LHC and the LC, or the pure jet mode at the LC \cite{lcllll}.
(Because of the large QCD background rate, the pure jet mode will be 
extremely difficult to utilize at hadron colliders.)

\section{ Backgrounds }
\indent\indent

For each decay mode of the $WW$ pair, the relevant backgrounds vary. But, 
in general, one of the dominant background processes is the {\it intrinsic}
electroweak background, which contains the same final state 
as the signal event.
This background rate in the TeV region can be generated by calculating
the Standard Model production rate of $f {\bar f} \ra f {\bar f} WW$ with a
light (e.g., 100\,GeV) SM Higgs boson \cite{llll}.
For example, the $W_LW_L$ signal rate in the TeV region from a 
1 TeV Higgs boson is equal to the difference between the event rates 
calculated using a 1 TeV Higgs boson and a 100 GeV Higgs boson \cite{llll}.

The other important backgrounds are:
the electroweak-QCD process $W+\,{\rm jets}$ 
(which contains a ``fake $W$'' mimicked by two QCD jets),  
and the $t \bar t$ pair (which subsequently decays to a $WW$ pair),
{\it etc}.
We now discuss these backgrounds in various $WW$ decay modes.
Without the loss of generality, in the rest of this article
we shall only consider the major background processes at the hadron 
collider LHC.
The backgrounds at the Linear Collider (LC) 
are usually easier to deal with 
because the initial state does not involve 
strong QCD interactions.

\subsection{ $Z^0(\ra \ell^+\ell^-)Z^0(\ra \ell^+\ell^-)$
and $Z^0(\ra \ell^+\ell^-)Z^0(\ra \nu \bar \nu)$ modes }
\indent\indent

The signature for the signal in this mode is either an event with 
four isolated leptons with high transverse momenta, 
or two isolated leptons associated with large missing transverse momentum
in the event.
The dominant background processes for this mode are
$q \bar q \ra Z^0 Z^0 X$, $g g \ra Z^0 Z^0 X$ \cite{llll},
where $X$ can be additional QCD jet(s).
The final state $Z^0$ pairs produced from the above processes 
tend to be transversely polarized.
Similarly, the $Z^0$ pairs produced from the 
{\it intrinsic} electroweak background process are also mostly 
transversely polarized.
This is because the coupling of a transverse $W$ boson 
to a light fermion (either quarks or leptons) is stronger than that of 
a longitudinal $W$ in the high energy region.
Hence, to discriminate the above backgrounds
from the signals, we have to study the polarization 
of the final state $W$ boson.
For the same reason, the gauge boson emitted from the initial state fermions
are likely to be transversely polarized.
To improve the ratio of signal to background rates, 
some kinematic cuts (to be discussed in the next section)
can be used to enhance the event sample in which the $W$ bosons
emitted from the incoming fermions are mostly longitudinally
polarized, and therefore can enhance the $\WWWW$ signal event.

Since it is easy to detect $Z^0 \ra \ell^+\ell^-$ with a good accuracy,
we do not expect backgrounds other than those discussed above to be large. 
Similarly, because of the  large
missing transverse momentum of the signal event,
the $Z^0 \ra \nu \bar \nu$ signature is difficult
to mimic by the other SM background processes.

\subsection{ $W^+(\ra \ell^+\nu) W^-(\ra \ell^- \bar \nu)$ mode}
\indent\indent

For this mode, in addition to the background processes
$q \bar q \ra q {\bar q} W^+W^-$, $q \bar q \ra W^+W^- X$ and
$g g \ra W^+W^- X$, the $t \bar t + \, {\rm jet}$ process
can also mimic the signal event because the 
final state top quark pair can decay into 
a $W^+W^-$ pair for the heavy 
top quark \cite{llll}.

\subsection{ $W^\pm (\ra \ell^\pm \nu) Z^0(\ra \ell^+ \ell^-)$   
and $W^\pm(\ra \ell^\pm \nu) W^\pm (\ra \ell^\pm \nu)$  modes}
\indent\indent

Besides the background processes similar to those discussed
above, the $Z^0 t \bar t$
event can also mimic the $W^\pm Z^0$ signal.

For the purely leptonic decay 
mode of $W^\pm W^\pm$ \cite{wpwpll}, the signature is two like-sign 
isolated leptons with high $P_T$ and large $\ETslash$.
There are no low-order backgrounds from quark-antiquark or gluon-gluon
fusion processes. However, other backgrounds can be important, such as
the production of the transversely polarized 
$W$-pairs from the {\it intrinsic} electroweak background 
process \cite{wpwpew} or from the QCD-gluon exchange 
process \cite{gluonex}, 
and the $W^\pm t \bar t$ production from the electroweak-QCD process
\cite{llll}.

\subsection{ $\WPWM$ mode }
\indent\indent

The dominant background processes for this mode are
$q \bar q \ra q {\bar q} W^+W^-$, $q \bar q \ra W^+W^- X$, 
and $g g \ra W^+W^- X$ \cite{qqww,ggww,wpwmew,wwjet}.
The signature for the signal in this mode is an 
isolated lepton with high transverse momentum $P_T$, 
a large missing transverse energy $\ETslash$, 
and two jets whose invariant mass is about the mass of the $W$-boson. 
The electroweak-QCD process $W^++ \, {\rm jets}$ 
 can mimic the signal when the invariant mass of the two 
QCD jets is around $M_W$ \cite{wtwoj,wthreej}.
Other potential background processes for this mode are the QCD processes
$ q \bar q, \, gg \ra t \bar t X $, $W t \bar b$ and 
$t {\bar t} + \, {\rm jet(s)}$ 
\cite{ttbaro,ttbarqcd,ttbar,wt,ttbarjet}, in which
a $W$ boson can come from the decay of $t$ or $\bar t$.

\subsection{ $\WPZ$ mode }
\indent\indent

The signature of the signal in this mode 
is an isolated lepton with high $P_T$, 
a large missing transverse energy $\ETslash$, 
and a two jet invariant mass around $M_Z$. 
The dominant background processes for this mode are similar to those for the 
$\WPWM$ mode discussed above. They are $ q_1 \bar q_2 \ra W^+ Z^0$,
$W^+Z^0 + \, {\rm jet(s)}$, $W^+ + \, {\rm jets}$ 
and $Z t \bar t$ production processes \cite{wwjet,wtwoj,wthreej,qqwz,zttbar}.

To separate this signal from $\WPWM$ a good jet energy resolution is needed
to distinguish the invariant mass of the two jets between $M_Z$ and $M_W$,
which differ by about 10 GeV.  Another technique to distinguish these
two kinds of events is to measure the average electric charge of the jets,
which has been applied successfully at LEP experiments \cite{aleph}.

As noted above, because of the large branching ratio,
the pure jet mode from the $W$ boson decay can also be useful
at the future lepton colliders \cite{lcllll}, where the dominant background 
for the detection of the $\WWWW$ signal event
is again the {\it intrinsic} electroweak process.

In general, without imposing any kinematic cuts,
the raw event rate of the signal is significantly
smaller than that of the backgrounds. 
However, the signature of the signal can actually be distinguished from that 
of the backgrounds so that some kinematic cuts can be applied
to suppress the backgrounds and enhance the signal-to-background ratio.
We shall examine the characteristic differences between 
the event structures of the signal and the backgrounds in the next section.

\section{ How to Distinguish Signal from Background }
\indent\indent

The signature of the signal event can be 
distinguished from that of the background events in 
many ways. We first discuss differences in the global features of the 
signal and the background events, then point out 
some distinct kinematics of the signal events.
To simplify our discussion, we shall only concentrate on the $\WPWM$ mode
in this section.

\subsection{ Global Features }
\indent\indent

The signal of interest is the $WW$ pair produced from the $W$-fusion
process. The spectator quark jet that emitted the
$W$-boson in the $W$-fusion 
process tends to go into the 
high rapidity region. This jet typically has a high 
energy, about 1 TeV, for $\MWW \sim 1$ TeV  ($\MWW$ is the invariant mass 
of the $WW$ pair.) Therefore, one can tag this forward jet to suppress 
backgrounds \cite{wwpt,tag,llll}.
As noted in the previous section, the $W$ boson pairs produced from the 
{\it intrinsic} electroweak process 
$ q \bar q \ra q \bar q W^+ W^-$ tend to be transversely polarized,
and the initial state gauge bosons are likely to be transversely 
polarized as well. 
To see how the forward jet can be used to discriminate the signal from 
the background events, we consider
the $W^+W^-$ fusion process as an example.
Since the coupling of the $W^\pm$ boson and the incoming quark is
purely left-handed, the out-going quark tends to go along with
the incoming quark direction when emitting a longitudinal 
$W$ boson, and opposite direction when emitting a transverse (left-handed) 
$W$. This can be easily understood from the helicity conservation. 
Hence, in the {\it intrinsic} background event, the out-going quark jet
is less forward (and less energetic) than that in the
signal event.

Furthermore, because the production mechanism of 
the signal event is purely electroweak, 
the charged particle multiplicity of the signal event is smaller than 
that of a typical electroweak-QCD process such as 
$q \bar q \ra g W^+W^-(\ra q_1 \bar q_2)$ 
or $qg \ra q W^+ q_1 \bar q_2$.
Because of the small hadronic activity in the signal event, in the central 
rapidity region there will be fewer hard QCD jets produced. 
At the parton level, they are the two quark jets produced from the $W$-boson 
decay plus soft gluon radiation.
 However, for the background process, such as $t \bar t$ production, 
there will be more than two hard jets 
in the central rapidity region both because 
of the additional jets from the decay of $t$ and $\bar t$ 
and because of the stronger hadronic 
activity from the effect of QCD color structure 
of the event.
Therefore, one can reject events with more than two 
hard jets produced in the central rapidity
region to suppress the backgrounds. This was first suggested in 
Ref.~\cite{ttbar} using a hadron level analysis to show how the $t \bar t$ 
background can be suppressed.

A similar trick of vetoing extra jets in the central rapidity region
was also applied at the parton level \cite{veto,llll}
for studying the pure leptonic decay mode of $W$'s.
An equivalent way of making use of the global 
difference in the hadronic activity of the events is to apply cuts on the 
number of charged particles. This was first pointed out in 
Refs.~\cite{multi} and \cite{kane}.
The same idea was later packaged in the context of 
selecting events with ``rapidity gap'' to enhance
the signal-to-background ratio \cite{gap}.

In the $W$-fusion process, the typical transverse momentum
of the final state $W$-pair is about $M_W/2$ \cite{wwpt}.
However, in the TeV region, the $P_T$ of the $W$-pair produced from the 
background process, such as $q \bar q \ra gWW$, can be 
of a few hundred GeV.
Therefore, the two $W$'s (either both real or one real and one fake)
produced in the background event are less back-to-back
in the transverse plane than those in the signal event.

\subsection{ Isolated Lepton in $W^+ \ra \ell^+ \nu$}
\indent\indent

Because the background event typically has more hadronic activity in the 
central rapidity region, 
the lepton produced from the $W$-boson decay is 
usually less isolated than that in the signal event.
 Therefore, requiring an isolated 
lepton with high $P_T$ is a useful method to suppress
the backgrounds. This requirement together with 
large missing transverse energy in the event 
insures the presence of a $W$-boson.
Finally, it is also important to be able to measure 
the sign of the lepton charge to reduce the backgrounds for the
detection of the $\WPWP$ mode \cite{llll}

\subsection { $W \ra q_1 \bar q_2$ decay mode}
\indent\indent

To identify the signal, we have to reconstruct the
two highest $P_T$ jets in the 
central rapidity region to obtain the invariant mass of the
$W$-boson. It has 
been shown \cite{kane}
 that an efficient way of finding these two jets is to first 
find a big cone jet with invariant mass around $M_W$, then demand that there 
are two jets with smaller cone size inside this big cone jet. 
Because we must measure any new activity in $\WWWW$, and
because the 
$W$-boson in the background event is mainly 
transversely polarized \cite{kane}, \footnote{
The same conclusion also holds for the QCD background event with 
``fake $W$'', which usually consists of one hard jet and one soft jet.
Hence, after boosting these two jets back into the rest frame of the
``fake $W$'', its angular distribution resembles that from a 
transverse $W$ boson.}
one must measure the fraction of longitudinally 
polarized $W$-bosons in the $WW$ pair data sample and compare with that 
predicted by various models of EWSB sector.
                    
It was shown in Ref. \cite{toppol} that a large fraction
($\sim 65\%$ for a 175 GeV top quark)
of the $W$ bosons from the top quark decays is longitudinally 
polarized.\footnote{
The ratio of the longitudinal versus the transverse $W$'s from top
quark decays is about $m_t^2/(2 M_W^2)$.}
This can in principle complicate the above method of using the 
fraction of longitudinally $W$ bosons to detect the signal.
Fortunately, after imposing the global cuts such as vetoing the central 
jet and tagging the forward jet, the $t \bar t$ backgrounds are small.
(If necessary, it can be further suppressed by vetoing event with $b$ jet,
because for every background event with top quark there is always a $b$ quark
from the SM top decay.)
To suppress the $W^+ t \bar b$ and $W^- {\bar t} b$ backgrounds 
\cite{wt}, which have smaller raw production rate than the $t \bar t$ event, 
the same tricks can be used. Furthermore, the top quark produced
in the $W^+ t \bar b$ event is mostly left-handed polarized
because the coupling of $t$-$b$-$W$ is purely left-handed in the 
SM. The kinematic cut of vetoing events with additional 
jets in the central rapidity region will reduce the fraction
of the events in which the $W$ boson from the top 
quark (with energy of the order 1 TeV) decay is longitudinally polarized.
This is because in the rest frame of
a left-handedly polarized top quark, 
which decays into a longitudinal $W$-boson, the decay
$b$-quark prefers to move along the moving direction of the top quark 
in the center-of-mass frame of the $W^+ t$ pair. Hence, such a
background event will produce an additional hard jet in the central 
rapidity region \cite{toppol}.

In the next section, we show how to observe the signals predicted by various 
models of the
symmetry breaking sector. Some of them were studied at the hadron 
level, some at the parton level. We shall  not reproduce those 
analyses but only sketch the ideas of various techniques used in detecting  
$\WWWW$ interactions. The procedures discussed here are not necessarily
the ones used in the analyses previously performed in the literature.
If the signal event rates are large enough to observe the purely
leptonic mode, then studying the symmetry breaking sector at the LHC 
shall be possible. However, a parton level study in Ref.~\cite{llll}
shows that the event rates are generally small after imposing the necessary
kinematic cuts to suppress the backgrounds. To clearly identify the
EWSB mechanism from the $\WWWW$ interactions, 
the $\ell^\pm + \, {\rm jet}$ mode of the $WW$ pair should also be studied
because of its larger branching ratio than the pure leptonic 
mode.\footnote{
At the LC, because its initial state is colorless, the pure jet decay
mode (with the largest branching ratio) of the $W$ boson can also
be useful.
}
That is the decay mode we shall concentrate on in the following section
for discussing the detection of various models of EWSB sector.

\section{ Various Models }
\indent\indent

\subsection{ A TeV Scalar Resonance }
\indent\indent

Based on the triviality argument \cite{trivial},
the mass of the SM Higgs boson cannot be much larger than 
$\sim 650$ GeV, otherwise the theory would be inconsistent. 
(If the SM is an effective theory valid up to the energy scale much 
higher than 1 TeV, then this number is even lower.)
However, one may consider an effective theory, such as an 
 electroweak chiral lagrangian, in which 
a TeV scalar (spin-0,isospin-0) resonance couples
to the would-be Goldstone bosons in the same way as the Higgs boson 
in the Standard Model \cite{chiralone,chiral,llll}.
(The mass and the width of the scalar resonance are the two free parameters 
in this model.)
Then one can ask how to detect such a TeV 
scalar resonance. This study was
already done at the hadron level in Ref.~\cite{kane}. 

The tricks of enhancing the ratio of signal 
to background are as follows.
First of all, we trigger on a
high $P_T$ lepton. The lepton is said to be isolated if there is no more than
a certain amount of hadronic energy inside a cone 
of size $\Delta R$ surrounding 
the lepton. ($\Delta R=\sqrt{(\Delta \phi)^2+(\Delta \eta)^2}$,
$\phi$ is the azimuthal angle and $\eta$ is the pseudo-rapidity.)
A TeV resonance produces a $W$-boson with typical $P_T$ at the order of 
$\sim 1/2$ TeV, therefore, the $P_T$ of the
lepton from the $W$-decay is at the order
of a few hundred GeV. 
The kinematic cut on the $P_T$ of an isolated lepton alone can
suppress a large fraction of $t \bar t$ background events because the lepton
produced from the decay of the $W$-boson typically has $P_T \sim m_t/3$, where
$m_t$ is the mass of the top quark. Furthermore, 
the lepton is also less isolated in
the $t \bar t$ event than that in the signal event.
After selecting the events with an
isolated lepton  with high $P_T$, we can make use of
the fact that the background event contains more hadronic activity than the
signal event to further suppress the background. One can make a cut on the
charged particle multiplicity of the event to enhance the 
signal-to-background ratio. 
The alternative way of making use of this fact is to
demand that there is only one big cone jet in the central rapidity region
of the detector \cite{kane}.
 The background process typically produces more hard jets
than the signal, hence vetoing the events with more than one big cone jet
in the central rapidity region is also a useful technique. 
The $W^++\,{\rm jets}$ and 
$t \bar t$ background
processes can further be suppressed by demanding that
the large cone jet has
invariant mass $\sim M_W$ and high $P_T$.
Inside this big cone jet, one requires two small cone 
jets corresponding to the two decay quark jets of the $W$-boson. 

As discussed above, measuring the polarization of the $W$ bosons in the 
final state can be a very useful tool for detecting and discriminating 
mechanisms of EWSB. Therefore, the best strategy for analyzing the
the experimental data is not to bias the information on the
polarization of the $W$ boson.
Some of the methods that can preserve the information on the
polarization of the $W$ boson were presented in Ref. \cite{kane}.
It was shown that it is possible to
measure the fraction of longitudinal $W$'s in the candidate $W$ samples to
distinguish various models of EWSB sector.
One of the techniques which would not bias the polarization of the $W$-boson
is to count the charged particle multiplicity inside the big cone jet. A real
$W$-boson decays into a color singlet state of $q \bar q$
with the same multiplicity regardless of its energy, hence the
charged particle multiplicity of these two jets is less than that of a pair
of non-singlet QCD jets (which form the
``fake $W$''), either quark or gluon jets. 
Furthermore, the QCD background events usually have more complicated 
color structure at the parton level, so that the hadron multiplicity of
the background event is generally larger than that of the signal event
in which the $WW$ system is a color singlet state.
Since the above methods only rely on the global features of the events,
they will not bias the information on the $W$ boson polarization.

Up to this point, we have only discussed the event structure in the central
rapidity region. As discussed in the previous section, in the large rapidity
region the signal event tends to have an energetic forward jet. It has been
shown that tagging one such forward jet can further suppress the background
at very little cost to the signal event rate \cite{llll}.

Furthermore, with rapidity coverage down to 5, one can have a good
measurement on the missing energy
($\ETslash$). Because 
the typical $\ETslash$ due to the neutrino
from the $W$-boson, with energy $\sim 1$\,TeV,
decay is of the order of a few hundred GeV, the
mis-measurement of neutrino transverse momentum 
due to the underlying hadronic activity is negligible.
 Knowing $\ETslash$ and the momentum of the lepton, one
can determine the longitudinal momentum of the
neutrino up to a two-fold solution
by constraining the invariant mass of the lepton and neutrino to
be $M_W$ \cite{kane}.
{}From the invariant mass of
 $\ell,\,\nu,\,q_1$, and $\bar q_2$, one can reconstruct 
$M_{WW}$ to discriminate background from signal events. 
If the width of the heavy resonance is small,\footnote{
For a SM Higgs boson with mass $m_H$ in units of TeV, its 
decay width would be about equal to $m_H^3/2$, which is not small.}
then one can detect a ``bump'' in the $M_{WW}$ distributions. 
However, if its width is too large, then the best way to detect 
this new physics effect is to measure the fraction ($f_L$) of longitudinal 
$W$'s in the event sample.

\subsection{ A TeV Vector Resonance }
\indent\indent

An example of this type of resonance is a techni-rho in the techni-color
model \cite{techni}.
 What we have in mind here
is a vector (spin-1,isospin-1) resonance in the electroweak chiral
lagrangian. The mass and the width of the vector resonance 
are the two free parameters
in this model. Because this resonance gives a large contribution in the 
isospin-1 channel, 
the most useful mode to look for such a resonance is the $W^\pm Z^0$ mode. 
If the signal event rate is large enough, the resonance can be observed by
the pure leptonic decay mode
$W^+(\ra \ell^+\nu)Z^0(\ra \ell^+\ell^-)$ in which all 
the leptons have
$P_T\sim$ few hundred GeV and are well isolated. If the $\WPZ$ mode is
necessary for the signal to be observed, the strategies discussed in the
previous subsection for the $\WPWM$ mode can be applied in this case as well.
Needless to say, in this case,
the invariant mass of the two jets peaks 
around $M_Z$ not $M_W$. It could be very valuable to improve the
techniques that separate $W( \ra jj)$ from $Z(\ra jj)$ by identifying
the average electric charge of each of the two decay jets.\footnote{
For the techniques used in identifying the average electric charge of 
a QCD jet, see, for example, Ref. \cite{aleph}.
}
 Obviously, the two jets from the $Z^0$ boson decay should have the same 
electric charges.

\subsection{ No Resonance }
\indent\indent

If there is no resonance at all,
the interactions among the longitudinal $W$'s become strong in the TeV region. 
Although the non-resonance scenario is 
among the most difficult cases to probe, 
this does {\it not} imply in any sense that 
it is less likely than the others to describe the underlying dynamics
of the electroweak symmetry breaking. 
For example, it was argued in Ref.~\cite{velt} that 
the non-resonance scenario may be likely to happen.
Within this non-resonance scenario, the electroweak chiral lagrangian (EWCL)
provides the most economic way to parameterize models of strongly coupled
EWSB sector.
The model with only the lowest order term 
(containing two derivatives) in the EWCL
is known as the low energy theorem model.
The signal of this model can be detected from studying the $\WPWM$ mode 
in the TeV region Ref.~\cite{kane}.
The techniques of observing this signal are identical to 
those discussed above.

In Ref. \cite{llll}, it was shown that it is possible to  
study the pure leptonic mode $\WPWP$ in
the multi-TeV region  to test 
the low energy theorem model as long as the integrated luminosity
 (or, the event rate) is large enough. 
The dominant backgrounds for this mode are 
the {\it intrinsic} background, $W^+ t \bar t$, and QCD-gluon
exchange processes.
The signal event can be triggered by two like-sign
charged leptons with high $P_T$ ($\sim $ few hundred GeV). One can 
further require
these leptons to be isolated and veto events with additional high $P_T$ jets in
the central rapidity region.
There are two missing neutrinos in the event 
so that it is difficult to reconstruct the $W$-boson and
measure its polarization. Hence,  in the
absence of a ``bump'' structure in any
distribution, one has to know the background event rate
well to study the EWSB sector,
unless the signal rate is very large.
Similarly, measuring the charged or total
particle multiplicity of the event and tagging
a forward jet can further improve the signal-to-background ratio.

Particularly for the case of no resonance, when the signal rate is not 
large, it is important to avoid imposing kinematic cuts which greatly
reduce the signal or  bias the polarization information 
of the $W$ bosons in the data sample.
The specific technique for measuring $f_L$, as proposed 
in Ref.~\cite{kane},  will probably have to be used to study the 
non-resonance case, and to probe the EWSB sector.
This technique takes advantage of the fact that the SM
is well tested, and will be much better tested in the 
TeV region by the time the study of $W_LW_L$ 
interactions is under way.
Every event of a real or fake $W_LW_L$ interaction will be
clearly identified as originating either from SM or new physics.
The real SM events 
(from $q \bar q , \, gg \ra WW$, $Wjj$, $t \bar t$, \etc)
can all be calculated and independently measured. 
Thus, one can first make global cuts such as requiring a high energy 
spectator jet and low total particle multiplicity,
and then examine all remaining candidate events 
to see if they are consistent with SM processes
or if they suggest new physics, in particular new sources of longitudinal
$W$'s. In principle, only one new quantity needs to be measured: the fraction
of $W_LW_L$ events compared to the total number of all $WW$ events 
including real and fake $W$'s. This can be done by the
usual approach of a maximum likelihood analysis, or
probably even better by the emerging neural network techniques
\cite{neural},
for which this analysis appears to be ideally suited.

Ultimately, recognizing that {\it in the TeV region}
every event must originate from either the
well understood Standard Model physics or beyond
 will be the most powerful approach to discovering
any deviations from the perturbative Standard Model predictions. 

\subsection{ Beam Pipe $W$'s }
\indent\indent

So far, we have only discussed signal events with high 
$P_T$ $W$-bosons produced in the central rapidity region. If there are many
inelastic channels opened in the $WW$ scattering 
process \cite{CG,inelast,nlf},
then based on the optical
theorem, the imaginary part of the forward elastic scattering amplitude is
related to the total cross section, and will not be
small \cite{nlf}. This
implies that it is possible for the final state $W$'s to
predominantly go down to
the beam pipe when produced from 
$W^+W^- \ra W^+W^-$ elastic scattering. 
Assuming this to be the case, it is
important to know how to detect such beam pipe $W$'s in the TeV region.

The typical transverse momentum of the 
decay particle in $W \ra f_1 f_2 $ is about $M_W/2$.
For $M_{WW}  > 2 M_W$,
the typical opening angle between the decay products 
of one of the $W$'s is about  ${4M_W / M_{WW}}$.  
Therefore, the absolute value of the rapidity 
of the decay products is likely 
to be within the range  2.5 to  4 for $M_{WW} \sim 1$ TeV.
With appropriate effort they should be detectable (perhaps not in
every detector, but certainly in some detectors eventually).
To suppress the backgrounds, one can veto
events with any jets or leptons in the central rapidity region, 
$|\eta| \le 2.5$.  
Another signature of the
signal event is the appearance of an energetic quark jet, 
the quark recoiling after emitting one of the interacting $W$'s,
with rapidity in the range 3 to 5.
One can thus further suppress QCD and electroweak backgrounds 
by tagging one forward (or backward) jet.
The background due to $W$'s emitted 
in a minimum bias event can also be suppressed, 
because, unlike the longitudinal $W$'s of the signal,
these $W$'s tend to be transversely polarized.
As a result, one of their decay products tends 
to be boosted more than the other, 
and is likely to be lost down the beam pipe, say, $|\eta| > 5$.  
Combining these techniques, 
we speculate that it may be feasible to detect 
longitudinal $W$ scattering
even in models in which $W$'s tend to be 
scattered predominantly along the beam pipe direction \cite{beam}.

\section{Sensitivities of High Energy Colliders to EWSB Sector}
\indent\indent

In the previous sections, we have discussed various methods suitable for 
detecting  the strongly coupled EWSB sector, of which 
a few models were briefly discussed as well.
Among them, the most difficult
one to detect is the no-resonance model, in which 
there is no resonance below the TeV scale.
Here, we shall consider this most difficult case and investigate 
the type of colliders and scattering processes needed to
completely probing the EWSB sector.

Below the scale of any new heavy resonance,
the EWSB sector can be parametrized by means of the
EWCL in which the 
$SU(2)_L \times U(1)_Y$ gauge symmetry is nonlinearly 
realized \cite{chiralone}.
Without experimental observation of any
new light resonance in the EWSB sector, this effective
field theory approach provides the most economic description of
the possible new physics effects 
and is thus {\it complementary} to those specific model buildings.
Hereafter, we shall concentrate on the effective bosonic operators, 
among which the leading order operators are model-independent,
while the other 15
next-to-leading-order (NLO) operators \cite{app}
depend on the details of models.
Furthermore, the 12 NLO operators
${\cal L}^{(2)\prime}$ and ${\cal L}_{1\sim 11}$
are $CP$-conserving,  and the 3 operators
${\cal L}_{12\sim 14}$ are $CP$-violating.
Among those operators which contribute to the quartic 
Goldstone boson interactions, the operators 
${\cal L}_{4,5}$  is $SU(2)_C$ invariant, while the operators
$~{\cal L}_{6,7,10}~$ violate the $SU(2)_C$ custodial symmetry.

Given the EWCL, we have to examine which are the colliders and processes
that should be used to sensitively probe the complete set of the
NLO operators. This was recently performed in Ref. \cite{et3}, 
in which two important techniques were used.
First, it was shown in Ref. \cite{et3} that the 
intrinsic connection between measuring the longitudinal 
weak-boson scatterings and probing the symmetry breaking sector
can be clearly formulated by noting the physical implication
of the electroweak Equivalence Theorem \cite{et3}. 
Second, based on this new formulation of the ET, it becomes 
straightforward to 
discriminate processes which are not sensitive to the EWSB by simply 
examining the high energy behavior of the 
physical S-matrices for the scattering processes using 
the generalized Weinberg's counting rules derived in
Ref.~\cite{powc}.
We note that Weinberg's counting rules were derived
for non-linear sigma model, in contrast, the 
 generalized Weinberg's counting rules were derived for electroweak 
chiral lagrangian which
is a gauge theory and contains not only the scalar fields 
but also the vector and the fermion fields.

The conclusion of Ref. \cite{powc} can be summarized in Table 1, in which
the leading contributions (~marked by $\surd~$) 
can be sensitively probed, while the sub-leading contributions
(~marked by $\triangle~$) can only be marginally sensitively 
probed. ($L/T$ denotes the longitudinal/transverse polarizations of
$~W^\pm ,~Z^0~$ bosons.) 
To save space, 
Table~1 does not contain those processes to which the NLO operators 
{\it only} contribute sub-leading amplitudes. Some
of these processes
are $~WW\rightarrow W\gamma ,Z\gamma +{\rm perm.}~$ 
and $~f\bar{f}^{(\prime )}\rightarrow W\gamma ,WW\gamma , WZ\gamma 
~$, which all have one external transverse $\gamma$-line and 
are at most marginally sensitive.
Further conclusions can be drawn as follows: \\
~~~{\bf (1).}
At LC(0.5), which is a LC with $\sqrt{S}=0.5$\,TeV, ${\cal L}_{2,3,9}$
can be sensitively probed via $e^-e^+ \ra W^-_L W^+_L$. \\ 
~~~{\bf (2).}
For $V_L V_L \ra V_L V_L$ scattering amplitudes, 
the model-dependent operators ${\cal L}_{4,5}$
and ${\cal L}_{6,7}$ can be probed 
most sensitively. 
 ${\cal L}_{10}$ can only be sensitively probed 
via the scattering process $Z_LZ_L \ra Z_LZ_L$ which 
is easier to detect at the LC(1.5) [a $e^-e^+$ or $e^-e^-$ collider 
with $\sqrt{S}=1.5$\,TeV] than at the LHC(14) [a pp collider with
$\sqrt{S}=14$\,TeV]. \\ 
~~~{\bf (3).}
The contributions from ${\cal L}^{(2)\prime}$~ and 
${\cal L}_{2,3,9}$ to the pure $4V_L$-scattering processes
 lose the $E$-power dependence by a 
factor of $2$. Hence, the pure $4V_L$-channel is 
not sensitive to these operators. 
[Note that ${\cal L}_{2,3,9}$  can be sensitively 
probed via $f {\bar f} \ra W_L^-W_L^+$ process at LC(0.5) and LHC(14).]
The pure $4V_L$-channel cannot probe ${\cal L}_{1,8,11\sim 14}$ which
can only be probed via processes with $V_T$('s). 
Among ${\cal L}_{1,8,11\sim 14}$,
the contributions from $~{\cal L}_{11,12}~$ to processes 
with $V_T$('s) are most important, although their contributions 
are relatively suppressed by a factor $gf_\pi /E$  as compared to
the leading contributions from 
${\cal L}_{4,5}$ to pure $4V_L$-scatterings.
Where the vacuum expectation value $f_\pi=246$\,GeV.
${\cal L}_{1,8,13,14}$ are generally suppressed by higher powers of
$gf_\pi /E$ and are thus the least sensitive.
The above conclusions hold for both LHC(14) 
and LC(1.5). \\
~~~{\bf (4).} 
At LHC(14), ${\cal L}_{11,12}$ 
can be sensitively probed via $q \bar q' \ra W^\pm Z$
whose final state is not electrically neutral. Since
this final state is not accessible at LC,
LC(0.5) is not be sensitive to these operators.
To sensitively probe ${\cal L}_{11,12}$ at LC(1.5), one has to measure
$e^-e^+ \ra W^-_L W^+_L Z_L$. 
\\ 
~~~{\bf (5).}
To sensitively probe ${\cal L}_{13,14}$,
a high energy $~e^-\gamma~$ linear collider 
%$\gamma \gamma(1.2)$ with $\sqrt{S}=1.2$\,TeV  
is needed for studying the processes 
$~e^-\gamma \ra \nu_e W^-_LZ_L,~e^-W^-_LW^+_L~$, 
in which the backgrounds \cite{eAback}~ are much
less severe than processes like $\gamma \gamma \ra W^+_L W^-_L$ at
a $\gamma\gamma$ collider~\cite{lcws93}.\footnote{The amplitude of
$\gamma \gamma \ra W^+_L W^-_L$ is of the order of 
$~e^2\frac{E^2}{\Lambda^2}~$, to which the operators ${\cal L}_{13,14}$
(and also ${\cal L}_{1,2,3,8,9}$) can contribute. Thus, this process
can be useful for probing ${\cal L}_{13,14}$ 
at a $\gamma\gamma$ collider after effectively suppressing its background.
}

We also note that to measure the individual coefficient of the 
NLO operator, one has to be able to separate , for example, the
$W^+W^- \ra Z^0 Z^0$ and the $Z^0Z^0 \ra Z^0 Z^0$ production processes.
Although this task can be easily done at the LC by detecting a forward
tagged lepton, it shall be a great challenge at the LHC because 
both the up- and down-type quarks from the initial state contribute to the 
scattering processes.
{}From the above conclusion,  we speculate\footnote{
To further reach a detailed quantitative conclusion, 
an elaborate and precise numerical study on all signal/background rates 
is necessary.}~ that 
if there is no new resonance below the TeV scale and 
the coefficients of the NLO 
operators are not much larger than that suggested by 
the naive dimensional analysis \cite{nda}, the LHC alone may not be 
able to sensitively measure all these operators,
and linear colliders are needed to {\it complementarily}
cover the rest of the NLO operators.
In fact, the different phases of 500 GeV and 1.5 TeV energies at the LC
are necessary because they will be sensitive to different 
NLO operators.  An electron-photon (or a photon-photon) 
collider is also useful for measuring
some of the NLO operators that distinguish models
of strongly coupled EWSB sector.
 
\section{ Discussions and Conclusions}
\indent\indent

If there is no light resonance present in the EWSB sector, the 
$\WWWW$ scatterings must become strong in the TeV region.
We have discussed how to experimentally study the 
strongly coupled electroweak symmetry 
breaking sector by observing
$\WWWW$ interactions in the TeV region,
emphasizing general features of the event structure in the 
signal and background events. Various techniques 
of enhancing the ratio of signal to background  
were also presented. We showed how to detect longitudinal $W$-bosons
either in the central rapidity region of the detector or in 
the beam pipe direction. 
We showed that it is possible to study the electroweak
symmetry breaking sector in the TeV region 
even when the $W_LW_L$ scattering is not resonant, as may be the most 
likely outcome \cite{velt}.
 However, to ensure a complete study of the symmetry breaking
sector, the beam
pipe $W$'s also need to be measured if no signal events
are found in the central rapidity region. 
In the previous section, we also discussed the sensitivities of the 
future high energy colliders, such as the LHC and the LC,
to probing the strongly coupled EWSB sector which is
parameterized by the NLO operators of the electroweak
chiral lagrangian.

Most of the proposals discussed here have been examined at the parton level
but not in detector simulations \cite{sdcgem}.
 They have been demonstrated to be 
promising techniques, but we cannot be sure they will work 
until the detector simulations are carried out by experimentalists.
Fortunately, there will be plenty of time to do those studies before the
data is available.

\vspace{0.3cm}
\noindent
{\bf Acknowledgments}

I would like to thank
J. Bagger, V. Barger, E. Berger, K. Cheung, S. Dawson, J. Gunion, T. Han, 
H.-J. He, G. Kane, R. Kauffman, Y.-P. Kuang,
G. Ladinsky, S. Mrenna, S.G. Naculich, F.E. Paige, R. Rosenfeld, 
J.L.~Rosner,
H.F.-W. Sadrozinski, A. Seiden, S. Willenbrock, A. J. Weinstein
and Y.-P. Yao for collaboration.
To them and 
U. Baur,  M.S. Chanowitz,  M. Einhorn, S.D. Ellis, 
G. Feldman, M.K. Gaillard, 
E.W.N. Glover, H. Haber, C. Kao, S. Meshkov,
F. Olness, L. H. Orr, W.W. Repko, M.E. Peskin, E. Poppitz,
 C.R. Schmidt, W.J. Stirling, 
Wu-Ki Tung, G. Valencia, H. Veltman, M. Veltman, R. Vega
and  E. Wang, I am grateful for discussions.
This work was funded in part by the NSF grant No. PHY-9507683.    

%\newpage

%%%%%% Beginning of Table 1. %%%%%%%%%%%%%%%%%%%%%%%%%%%%%%%%%%%%%%%%%%%%%%%%%%
\newpage
%\addtolength{\textwidth} {0.8in}
%\addtolength{\oddsidemargin} {-0.8in}
%\addtolength{\evensidemargin}{-0.8in}

\tabcolsep 1pt
\begin{table}[H]  
\begin{center}

{\bf Table. 1}
~Probing the EWSB Sector at High Energy Colliders: \\
$\qquad\qquad\qquad$ 
A Global Classification for the NLO Bosonic Operators   \\
\vspace{0.3cm} 
\small (~Notations: ~$\surd =~$Leading contributions, 
$~\triangle =~$Sub-leading contributions,~ 
and ~$\bot =~$Low-energy contributions.~) \\ 
(~Notes:~ $^{\dagger}$Here, $~{\cal L}_{13}$ or $~{\cal L}_{14}~$ 
does not contribute at this order. \\    
$\qquad\qquad\qquad$ $^\ddagger$At LHC($14$),  
$W^+W^+\ra W^+W^+$ should also be included.~)  

\vspace{0.4cm}

\small

%%%%%{tab:smtab}

\small
%\begin{tabular}{||c||c|c|c|c|c|c|c|c|c|c||c||} 
%\begin{tabular}{||c||c|c|c|c|c|c|c|c|c|c||c||} 
\begin{tabular}{c|cccccccccc|c} 
%\begin{tabular}{||c|cccccccccc|c||} 
\hline\hline
& & & & & & & & & & &  \\
 Operators 
& $ {\cal L}^{(2)\prime} $ 
& $ {\cal L}_{1,13} $ 
& $ {\cal L}_2 $
& $ {\cal L}_3 $
& $ {\cal L}_{4,5} $
& $ {\cal L}_{6,7} $ 
& $ {\cal L}_{8,14} $ 
& $ {\cal L}_{9} $
& $ {\cal L}_{10} $
& $ {\cal L}_{11,12} $
& Processes \\
& & & & & & & & & & &  \\
\hline\hline
 LEP-I (S,T,U) 
& $\bot$ 
& $\bot~^\dagger$
&  
& 
& 
& 
& $\bot~^\dagger$
& 
&
&
& $e^-e^+\ra Z \ra f\bar{f}$\\ 
\hline
  LEP-II
& $\bot$ 
& $\bot$  
& $\bot$  
& $\bot$  
&  
& 
& $\bot$  
& $\bot$ 
&
& $\bot$  
& $e^-e^+ \ra W^-W^+$\\
\hline
  LC($0.5$)/LHC($14$)
& 
& 
& $\surd$
& $\surd$
& 
& 
& 
& $\surd$
&
& 
& $f \bar f\ra W^-W^+ /(LL)$\\  
& 
& $\triangle$
& $\triangle$
& $\triangle$
& 
& 
& $\triangle$
& $\triangle$
&
& $\triangle$
& $f \bar f\ra W^-W^+/(LT) $\\  
\hline
& 
& 
& 
& $\surd$
& $\surd$
& $\surd$
& 
& $\surd$
&
& $\surd$
& $f \bar f\ra W^-W^+Z /(LLL) $\\
& 
& $\triangle$ 
& $\triangle$
& $\triangle$
& $\triangle$
& $\triangle$
& $\triangle$
& $\triangle$
&
& $\triangle$
& $f \bar f\ra W^- W^+ Z /(LLT)  $\\
& 
& 
&  
& $\surd$
& $\surd$
& $\surd$
& 
& 
& $\surd$
& 
& $f \bar f \ra ZZZ /(LLL) $\\
& 
& 
& 
& 
& $\triangle$
& $\triangle$
& 
& 
& $\triangle$
& 
& $f \bar f \ra ZZZ  /(LLT)  $\\
 ~LC($1.5$)/LHC($14$)~ 
& 
& 
& 
& 
& $\surd$
& 
& 
&
& 
& 
& $W^-W^- \ra W^-W^- /(LLLL)~^\ddagger$\\
&
& 
& 
& $\triangle$
& $\triangle$
& 
& 
& $\triangle$
&
& $\triangle$
& $W^-W^-\ra W^-W^- /(LLLT)~^\ddagger$ \\
& 
& 
& 
& 
& $\surd$
& $\surd$
&
& 
&
& 
& $W^-W^+ \ra ZZ ~\&~{\rm perm.}/(LLLL)$ \\
&
& 
& $\triangle$
& $\triangle$
& $\triangle$
& $\triangle$
& 
& $\triangle$
&
& $\triangle$
& $W^-W^+ \ra ZZ ~\&~{\rm perm.} /(LLLT)$ \\
& 
& 
& 
& 
& $\surd$
& $\surd$
& 
& 
& $\surd$ 
&  
& $ZZ\ra ZZ /(LLLL) $\\
&
& 
& 
& $\triangle$
& $\triangle$
& $\triangle$
&  
&
& $\triangle$
&
& $ZZ\ra ZZ /(LLLT) $\\
\hline
& 
& 
& 
& $\surd$
& 
& 
& 
&
& 
& $\surd$
& $q\bar{q'}\ra W^\pm Z /(LL) $\\
& 
& $\triangle$
& $\triangle$
& $\triangle$
& 
& 
& $\triangle$
& $\triangle$
&
& $\triangle$
& $q\bar{q'}\ra W^\pm Z /(LT) $\\
 LHC($14$)
& 
& 
& 
& $\surd$
& $\surd$
& 
& 
& $\surd$
&
& $\surd$
& $q \bar{q'}\ra W^-W^+W^\pm /(LLL) $\\
& 
& 
& $\triangle$
& $\triangle$
& $\triangle$
&
& $\triangle$
& $\triangle$
&
& $\triangle$
& $q \bar{q'} \ra W^- W^+W^\pm  /(LLT)  $\\
& 
& 
& 
& $\surd$
& $\surd$
& $\surd$
& 
& 
&
& $\surd$
& $q \bar{q'}\ra W^\pm ZZ /(LLL) $\\
& 
& $\triangle$
& $\triangle$
& $\triangle$
& $\triangle$
& $\triangle$
& $\triangle$
& $\triangle$
&
& $\triangle$
& $q \bar{q'} \ra W^\pm ZZ  /(LLT)  $\\
\hline
 LC($e^-\gamma$)
& 
& $\surd$
& $\surd$
& $\surd$
& 
& 
& $\surd$
& $\surd$
&
& $\surd$
& $e^-\gamma \ra \nu_e W^-Z /(LL)$\\
& 
& 
& 
& 
& 
& 
&
&
&
&
& $e^-\gamma \ra e^-W^-W^+ /(LL)$\\
\hline
 LC($\gamma \gamma$)
& 
& $\surd$
& $\surd$
& $\surd$
& 
& 
& $\surd$
& $\surd$
&
& 
& $\gamma \gamma \ra W^-W^+ /(LL)$\\
& & & & & & & & & & &  \\
\hline\hline 
\end{tabular}

\end{center}
\end{table}
%%%%%%%%%%%% End of Table 1. %%%%%%%%%%%%%%%%%%%%%%%%%%%%%%%%%%%%%%%%%%%%%%%%  

\end{document}